\newcommand{\pg}{\ifmmode{{\rm PG}\,1416-129} \else PG\,1416--129\fi} 
\footnotesep \baselineskip	
\newenvironment{references}
{\normalbaselineskip=8pt \parskip 0pt \parindent=0pt
\everypar={\hangafter=1 \hangindent=3.5em \parindent=0pt \relax}
\centerline{\bf References}
\bigskip
}{}
{ \normalbaselineskip=8pt \parskip 0pt \parindent=0pt
\everypar={\hangafter=1 \hangindent=3.5em \parindent=0pt \relax}
\small
{\bf figure~ #1\/:}
}
{}
\newcommand{\ax}{\ifmmode{\alpha_x} \else $\alpha_x$\fi} 
\newcommand{\aE}{\ifmmode{\alpha_E} \else $\alpha_E$\fi} 
\newcommand{\atoms}{\ifmmode{\rm ~atoms~cm^{-2}} \else ~atoms cm$^{-2}$\fi}
\newcommand{\nh}{\ifmmode{\rm N_{H}} \else N$_{H}$\fi}
\newcommand{\nhgal}{\ifmmode{ N_{H}^{Gal}} \else N$_{H}^{Gal}$\fi}
\newcommand{\nhintr}{\ifmmode{ N_{H}^{intr}} \else N$_{H}^{intr}$\fi}
\newcommand{\nhtot}{\ifmmode{ N_{H}^{tot}} \else N$_{H}^{tot}$\fi}
\newcommand{\meangamma}{\ifmmode{\langle\Gamma\rangle} \else $\langle\Gamma\rangle$\fi}
\newcommand{\lopt}{\ifmmode l_{opt} \else $~l_{opt}$\fi}
\newcommand{\loglopt}{\ifmmode{\rm log}~l_{opt} \else log$~l_{opt}$\fi}
\newcommand{\lx}{\ifmmode l_x \else $~l_x$\fi}
\newcommand{\loglx}{\ifmmode{\rm log}~l_x \else log$~l_x$\fi}
\newcommand{\aox}{\ifmmode{\alpha_{ox}} \else $\alpha_{ox}$\fi} 
\newcommand{\auv}{\ifmmode{\alpha_{uv}} \else $\alpha_{uv}$\fi} 
\newcommand{\auvx}{\ifmmode{\alpha_{uvx}} \else $\alpha_{uvx}$\fi} 
\newcommand{\luv}{\ifmmode l_{uv} \else $~l_{uv}$\fi}
\newcommand{\logluv}{\ifmmode{\rm log}\,l_{uv} \else log$\,l_{uv}$\fi}
\newcommand{\ein}{{\em Einstein{\rm}}}

\newcommand{\ros}{{\em ROSAT{\rm}}}

\newcommand{\IUE}{{\em IUE}}
\newcommand{\HST}{{\em HST}}
\newcommand{\nufnu}{\ifmmode \nu f_{\nu} \else$\nu f_{\nu}$\fi}
\newcommand{\fnu}{\ifmmode f_{\nu} \else$f_{\nu}$\fi}

\newcommand{\mdot}{\ifmmode {\dot{M}}\else${\dot{M}}$\fi}
\newcommand{\rr}{\ifmmode {F_{PL} \over F_{disk}}\else${F_{PL} \over F_{disk}}$\fi}
\newcommand{\costheta}{\ifmmode {\cos \theta}\else${\cos \theta}$\fi}
\newcommand{\astar}{\ifmmode {a_{*}}\else${a_{*}}$\fi}

\newcommand{\micron}{\ifmmode\mu{\rm m}\else$\mu{\rm m}$\fi}
\newcommand{\fcgs}{\ifmmode erg~cm^{-2}~s^{-1[B}\else erg~cm$^{-2}$~s$^{-1}$\fi}
\newcommand{\lcgs}{\ifmmode erg~~s^{-1[B}\else erg~s$^{-1}$\fi}
\newcommand{\fnucgs}{\ifmmode erg~cm^{-2}~s^{-1}~Hz^{-1}\else erg~cm$^{-2}$~s$^{-1}$~Hz$^{-1}$\fi}
\newcommand{\lnucgs}{\ifmmode erg~s^{-1}~Hz^{-1}\else erg~s$^{-1}$~Hz$^{-1}$\fi}
\newcommand{\kms}{\ifmmode~{\rm km~s}^{-1}\else ~km~s$^{-1}~$\fi}
\newcommand{\mone}{\ifmmode ^{-1}\else$^{-1}$\fi}
\newcommand{\mtwo}{\ifmmode ^{-2}\else$^{-2}$\fi}
\newcommand{\arcsec}{\ifmmode ^{\prime\prime}\else$^{\prime\prime}$\fi}
\newcommand{\arcmin}{\ifmmode ^{\prime}\else$^{\prime}$\fi}
\newcommand{\degs}{\ifmmode ^{\circ}\else$^{\circ}$\fi}
\newcommand{\mv}{\ifmmode {m_{V}}\else${m_{V}}$\fi}
\newcommand{\Mv}{\ifmmode {M_{V}}\else${M_{V}}$\fi}
\newcommand{\msun}{\ifmmode {M_{\odot}}\else${M_{\odot}}$\fi}
\newcommand{\rsun}{\ifmmode {R_{\odot}}\else${R_{\odot}}$\fi}
\newcommand{\lsun}{\ifmmode {L_{\odot}}\else${L_{\odot}}$\fi}
\newcommand{\ciii}{\ifmmode{{\rm C\,III]}} \else C\,III]\fi}
\newcommand{\civ}{\ifmmode{{\rm C\,IV}} \else C\,IV\fi}
\newcommand{\heii}{\ifmmode{{\rm He\,II}} \else He\,II\fi}
\newcommand{\lya}{\ifmmode{{\rm Ly}\alpha}\else Ly$\alpha$\fi}
\newcommand\lyb{\ifmmode {\rm Ly}\beta \else Ly$\beta$\fi}
\newcommand{\mgii}{\ifmmode{{\rm Mg\,II}} \else Mg\,II\fi}
\newcommand{\ovi}{\ifmmode{{\rm O\,VI}} \else O\,VI\fi}
\newcommand{\siiv}{\ifmmode{{\rm Si\,IV}} \else S\,IV\fi}
\newcommand{\oiii}{\ifmmode{{\rm [O\,III]}} \else [O\,III]\fi}
\newcommand{\ew}{\ifmmode{W_{\lambda}} \else $W_{\lambda}$\fi}
\newcommand{\wciii}{\ifmmode{W_{\lambda}({\rm C\,III]})} \else $W_{\lambda}$(C\,III])\fi}
\newcommand{\wciv}{\ifmmode{W_{\lambda}({\rm C\,IV})} \else $W_{\lambda}$(C\,IV)\fi}
\newcommand{\wheii}{\ifmmode{W_{\lambda}({\rm He\,II})} \else $W_{\lambda}$(He\,II)\fi}
\newcommand{\wlya}{\ifmmode{W_{\lambda}({\rm Ly}\alpha)}\else $W_{\lambda}$(Ly$\alpha$)\fi}
\newcommand\wlyb{\ifmmode{ W_{\lambda}({\rm Ly}\beta )} \else $W_{\lambda}$(Ly$\beta$)\fi}
\newcommand{\wmgii}{\ifmmode{W_{\lambda}({\rm Mg\,II})} \else $W_{\lambda}$(Mg\,II)\fi}
\newcommand{\wovi}{\ifmmode{W_{\lambda}({\rm O\,VI})} \else $W_{\lambda}$(O\,VI)\fi}
\newcommand{\wsiiv}{\ifmmode{W_{\lambda}({\rm Si\,IV})} \else $W_{\lambda}$(Si\,IV)\fi}
\newcommand{\woiii}{\ifmmode{{W_{\lambda}(\rm [O\,III])}} \else \$W_{\lambda}($[O\,III])\fi} 
\newcommand{\rciii}{\ifmmode{{\rm C\,III]/Ly} \alpha} \else C\,III]/Ly$\alpha$\fi}
\newcommand{\rciiiciv}{\ifmmode{{\rm C\,III]/C\,IV}} \else C\,III]/C\,IV\fi}
\newcommand{\rciv}{\ifmmode{{\rm C\,IV/Ly} \alpha} \else C\,IV/Ly$\alpha$\fi}
\newcommand{\rheii}{\ifmmode{{\rm He\,II/Ly} \alpha} \else He\,II/Ly$\alpha$\fi}
\newcommand{\rmgii}{\ifmmode{{\rm Mg\,II/Ly} \alpha} \else Mg\,II/Ly$\alpha$\fi}
\newcommand{\rovi}{\ifmmode{{\rm O\,VI/Ly} \alpha} \else O\,VI/Ly$\alpha$\fi}
\newcommand{\roviciv}{\ifmmode{{\rm O\,VI/C\,IV}} \else O\,VI/C\,IV\fi}
\newcommand{\rsiiv}{\ifmmode{{\rm Si\,IV/Ly} \alpha} \else Si\,IV/Ly$\alpha$\fi}
\newcommand{\roiii}{\ifmmode{{\rm [O\,III]/H} \beta} \else [O\,III]/H$\beta$\fi}
\documentstyle[12pt,aaspp4]{article}

\received{ApJ, Dec 11, 1996}
\accepted{Feb 14, 1997}

\lefthead{Green et al.}
\righthead{Evidence Against BALS in PG\,1416--129}

\begin{document}
\title{Evidence Against BALS in the X-ray Bright QSO PG\,1416--129}

\author{Paul J. Green\altaffilmark{1}, Thomas
L. Aldcroft\altaffilmark{2}, Smita Mathur\altaffilmark{3}} 
\affil{Harvard-Smithsonian Center for
Astrophysics, 60 Garden St., Cambridge, MA 02138}
\altaffiltext{1}{pgreen@cfa.harvard.edu}
\altaffiltext{2}{taldcroft@cfa.harvard.edu}
\altaffiltext{3}{smathur@cfa.harvard.edu}

\and

\author{Norbert Schartel\altaffilmark{4}}
\affil{ESA, IUE Observatory, P.O. Box 50727, E-28080 Madrid,
Spain. Affiliated to the Astrophysics Division, Space Science
Department, ESTEC}
\altaffiltext{4}{nrs@encina.vilspa.esa.es}

\begin{abstract}
Recent results from the \ros\, All Sky Survey, and from deep \ros\,
pointings reveal that broad absorption line quasars (BALQSOs) are weak
in the soft X-ray bandpass ($\aox>1.8$) in comparison to QSOs with
normal OUV spectra ($\overline{\aox}=1.4$).  One glaring
exception appeared to be the nearby BALQSO
\pg, which is a bright \ros\, source showing no evidence for intrinsic 
soft X-ray absorption.  We present here our new HST FOS spectrum of
\pg, in which we find no evidence for BALs.  We show that the features
resulting in the original BAL classification, based on $\IUE$\, spectra,
were probably spurious.  On the basis of UV, X-ray and optical evidence,
we conclude that \pg\, is not now, and has never been a BALQSO.  Our
result suggests that weak soft 
X-ray emission is a defining characteristic of true BALQSOs.  If
BALQSOs indeed harbor normal intrinsic spectral energy distributions,
their observed soft X-ray weakness is most likely the result of
absorption.  The ubiquitous occurrence of weak soft X-ray emission with
UV absorption (BALs) thus suggests absorbers in each energy
regime that are physically associated, if not identical.
\end{abstract}


\section{INTRODUCTION}
\label{intro}

Broad absorption lines (BALs) are seen in about 10 - 15\% of
optically-selected QSOs, only among the radio-quiet (RQ) QSO population
(Stocke et al. 1992).  The optical/UV spectra show deep, wide
absorption troughs, displaced to the blue of their corresponding
emission lines (most often in the high ionization transitions of
C\,IV, Si\,IV, N\,V, and O\,VI), which are 
suspected to result from a line of sight passing through highly
ionized, high column density absorbing clouds outside the broad
emission line region (BELR). These clouds flow outward from the
nuclear region at speeds up to 0.1 - 0.2c.  Low BAL cloud covering
factors and the absence of emission lines at the high velocities observed
in BALQSOs, along with the similarity of emission-line and continuum
properties of BAL and non-BALQSOs (Hamann, Korista, \& Morris 1993,
Weymann et al. 1991) suggest that
{\em all} RQ QSOs (which in turn comprise $\sim 90\%$ of all
QSOs) have BAL clouds.  Thus BALQSOs are by no means exotic, but
rather represent a privileged line of sight toward the AGN nucleus
that probes clouds that are very near, or cospatial with, the BELR.

The absorbing columns (e.g., $N_H\sim 10^{19}$ to $10^{20}$) inferred
for BAL clouds from the OUV data (Hamann et al. 1993,
Turnshek 1988) are such 
that {\it a priori} one expects very {\em little} soft X-ray
absorption ($\tau<<1$).  However, Green et al. (1995) and Green \&
Mathur (1996; hereafter GM96) recently demonstrated that, when
compared to normal RQ QSOs, BALQSOs are weak in the soft X-ray
bandpass.  If BALQSOs harbor normal intrinsic
spectral energy distributions (SEDs), their soft X-ray weakness
is most likely the result of absorption.  Although this
remains to be proven for BALQSOs as a class, strong X-ray absorption
of a normal powerlaw continuum is clearly observed in the ASCA
spectrum of the prototype BALQSO PHL\,5200 ($N_H^{intr} \sim10^{23}$;
Mathur, Elvis, \& Singh 1995).   

The Green et al. (1995) sample of 36 BALQSOs was chosen from a large
uniformly-selected QSO sample (the LBQS), as observed during the \ros\,
All-Sky Survey (RASS). Although the short ($\sim 600$~sec) exposure times of
the RASS meant that the upper limits (for 35 of the 36 QSOs) were not
very sensitive, by stacking the X-ray data, they were able to show
that their uniform BALQSO sample was X-ray quiet at the 99.5\% 
significance level compared to carefully chosen comparison RQ QSO samples.  

  Then using deep pointed observations from the \ros~ PSPC, GM96
confirmed that BALQSOs are weak in the soft X-ray bandpass in
comparison to RQ QSOs with normal OUV spectra.  Nine out of twelve
reputed BALQSOs in their sample were not detected by \ros, the deep
pointings generally yielding $\aox>1.8$.  A comparison sample of 10
similar RQ QSOs (from Laor et al. 1994) without BALs yielded a sample mean 
\footnote[1]{The value of \aox\, is known to increase with \lopt\,
(Wilkes et al. 1994, Avni et al. 1995; Green et al. 1995).
However, the difference in \aox\, between the Laor et al. (1994)
sample and the GM96 BALs is much more than can be attributed to the
difference in their mean optical luminosities  (see discussion in
\S~\ref{discussion}).} of $\overline{\aox} =1.45\pm0.08$. 
If indeed the central continuum 
source of BALQSOs is similar to that of other QSOs, as argued above,
the intrinsic absorbing columns required to explain the observed soft
X-ray deficit ($N_{H}^{intr}>2\times10^{22}$cm$^{-2}$) must be at
least 100 times higher than those inferred from the UV data alone. In
contrast, the non-BAL sample shows no evidence at all for absorption.
Of the three remaining BALQSOs in GM96, 0312--555 was too distant for an
interesting lower limit to \aox. In a 58ksec summed exposure, the BALQSO
1246--057 was detected, yielding $\aox=1.98$, and $N_{H}^{intr}\sim
10^{23}$cm$^{-2}$.  Only one QSO, \pg, was X-ray bright, with a value
of $\aox=1.4$ typical of non-BALQSOs.

That all BALQSOs but \pg\, have large \aox\, indeed suggests 
some physical connection between the UV and X-ray absorption, assuming
similar intrinsic SEDs.  However, if PG1416--129 is a {\em bona fide}
BALQSO, although it would be uniquely 
well-suited for observational tests of absorber models, it would refute
the hypothesis that highly ionized UV absorbers in BALQSOs are also
responsible for their X-ray silence.  If instead a high quality UV
spectrum reveals only {\em associated} absorption (narrow optical and
ultraviolet absorption lines within the profiles of their broad
emission lines), that model may stand.  There are strong hints
that a continuous distribution of UV/X-ray absorbing properties may
exist between `associated absorbers' and BALs; PG1416--129 might just
be a 'missing link'. 

Techniques have recently been developed that simultaneously exploit UV
and X-ray spectra of QSOs with narrower line, associated absorbers 
to constrain the allowed ranges of the absorbing
cloud conditions through detailed photoionization modeling 
(e.g., 3C351, 3C212, NGC5548, NGC3516; Mathur 1994, Mathur et
al. 1994, 1995, 1996).  Although these UV/X-ray techniques have shown
that {\em consistent} physical conditions for both the UV and soft
X-ray absorbers can in many instances be derived, there is still
debate on whether they are physically associated (e.g., Kriss et
al. 1996a, b).  An application of these techniques to 
BALQSOs may eventually provide stronger constraints on BAL clouds, but
the weakness of BALQSOs in the soft X-ray regime makes this a
daunting task.  

  If \pg\, is a true BALQSO, its relative X-ray brightness and
proximity to earth (\pg, at $z=0.129$, has a redshift lower than any
confirmed BALQSO) would provide a uniquely accessible object for
detailed X-ray/UV studies of high column density absorbers near the
central engine of a QSO.  At the same time,
\pg\, provides a litmus test for the hypothesis that BALQSOs are
X-ray quiet as a class.  We were thus led to examine its UV spectrum
more carefully, as described below: is \pg\, a true BALQSO?

\vskip0.2cm
\noindent
\section{\IUE\, Spectra of PG1416--129}

  The BAL classification for \pg\, was originally awarded by Turnshek
\& Grillmair (1986), based on a single \IUE\, spectrum
(SWP8916), and propagated in a number of subsequent papers (e.g.,
Ulrich 1988, deKool \& Meurs 1994, Staubert, R. \& Maisack 1996).  That 
spectrum appears to show some evidence for what might be broad
absorption in any of 
CIV, SiIV, or Ly$\alpha$.  Blueward of CIV in particular, there are possible
absorption troughs that span $>2000$~km~s\mone, extending as far as about
20,000~km~s\mone\, from the line center.  However, the combination of SiIV
broad emission, with a spurious flux spike near 1665\AA\, might merely combine
to give that impression.  In addition, a narrow absorption trough
appears to bisect the CIV emission just redward of the line 
core, but no similar absorption is seen in \lya.

Since then, three other \IUE\, spectra have been obtained.  A log of
these observations is presented, along with continuum flux and \ew\,
measurements for the CIV emission line, in Table~\ref{tuv}.  We
believe that none of these spectra show BALs.  Neither does
an average spectrum, whether weighted by signal-to-noise ratio 
(SNR) or not (e.g., see the optimally-extracted and co-added spectrum 
of PG1416$-$129 from Lanzetta, Turnshek, \& Sandoval 1993).
However, an SNR-weighted sum of the \IUE\, spectra is dominated
by SWP33030 (Fig. 2), which shows features blueward of CIV
that could only optimistically be interpreted as BALs.
An examination of reference spectra showing camera artifacts in \IUE\,
SWP spectra (Crenshaw et al. 1990) is revealing.  Although the
strengths, both relative and absolute, of camera artifacts can vary,
the strongest spike for point sources is generally at 1663\AA,
blueward of the CIV emission line in \pg.  Other spurious
features, like those near 1700\AA\, probably contributed to the
original BAL classification.

 The unusual (and possibly variable) nature of the putative BALs 
in \pg, together with its X-ray brightness, led us to seek another 
UV spectrum, this time using the Hubble Space Telescope (HST).  
None of the artifacts just discussed are visible in our new HST 
spectrum, nor are any features reminiscent of BALs.

\section{HST FOS Observations}

On UT 23 August 1996, we obtained a 1-orbit (940sec) spectrum of \pg\,
with the Faint Object Spectrograph (FOS) on the Hubble Space
Telescope, using a 0.43\arcsec aperture and the G190H grating with the
blue detector.  We are aware of the scattered light problem in the FOS
when observing at the shortest wavelengths (Rosa 1994).  In our case, we
would not expect a significant scattered light component because
\pg\, has a typical blue power-law continuum.  Nevertheless, it is
prudent to make certain that the FOS spectrum does not approach zero
intensity bluewards of CIV, as it appears to do in the spectrum of
Turnshek \& Grillmair (1986).  We used the BSPEC program (Rosa 1994)
to simulate the scattered red light in the FOS G190H bandpass.  The
input spectral energy distribution was derived from an optical
spectrum of \pg\, (kindly provided by B. Wilkes) which covers the
range 3200 - 6000\,\AA.  We find a negligible contribution from
scattered light.

\begin{figure} 
\plotone{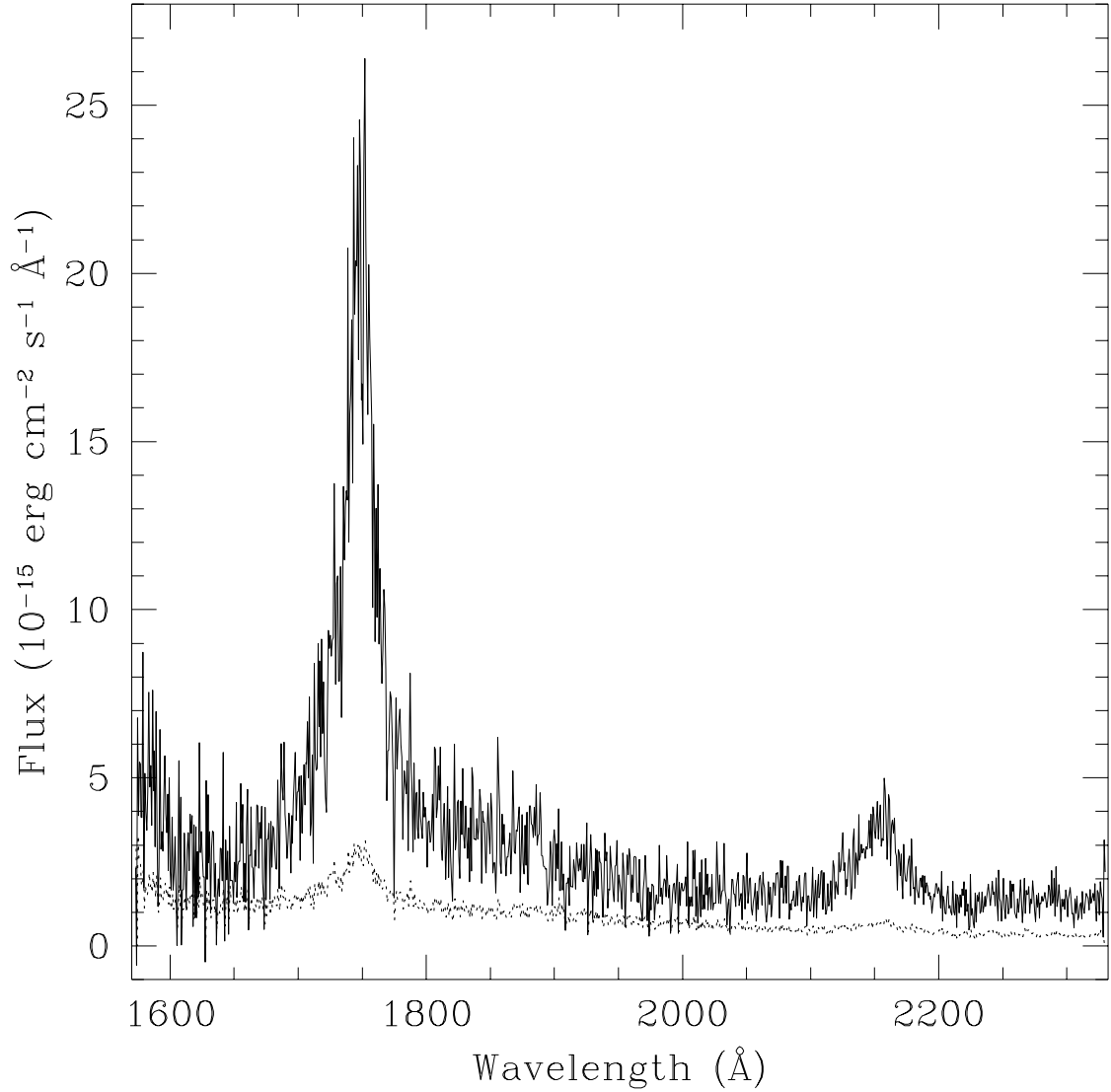}
\vfill
\caption{An HST FOS spectrum of \pg\, from 23 Aug 1996 (solid line),
block averaged by two, with its associated error (dashed
line). No evidence for either broad or narrow associated absorption 
is seen in the HST spectrum at CIV (1749\AA\, in the observed frame).  
}
\label{fullspec}
\end{figure}

\begin{figure} 
\plotone{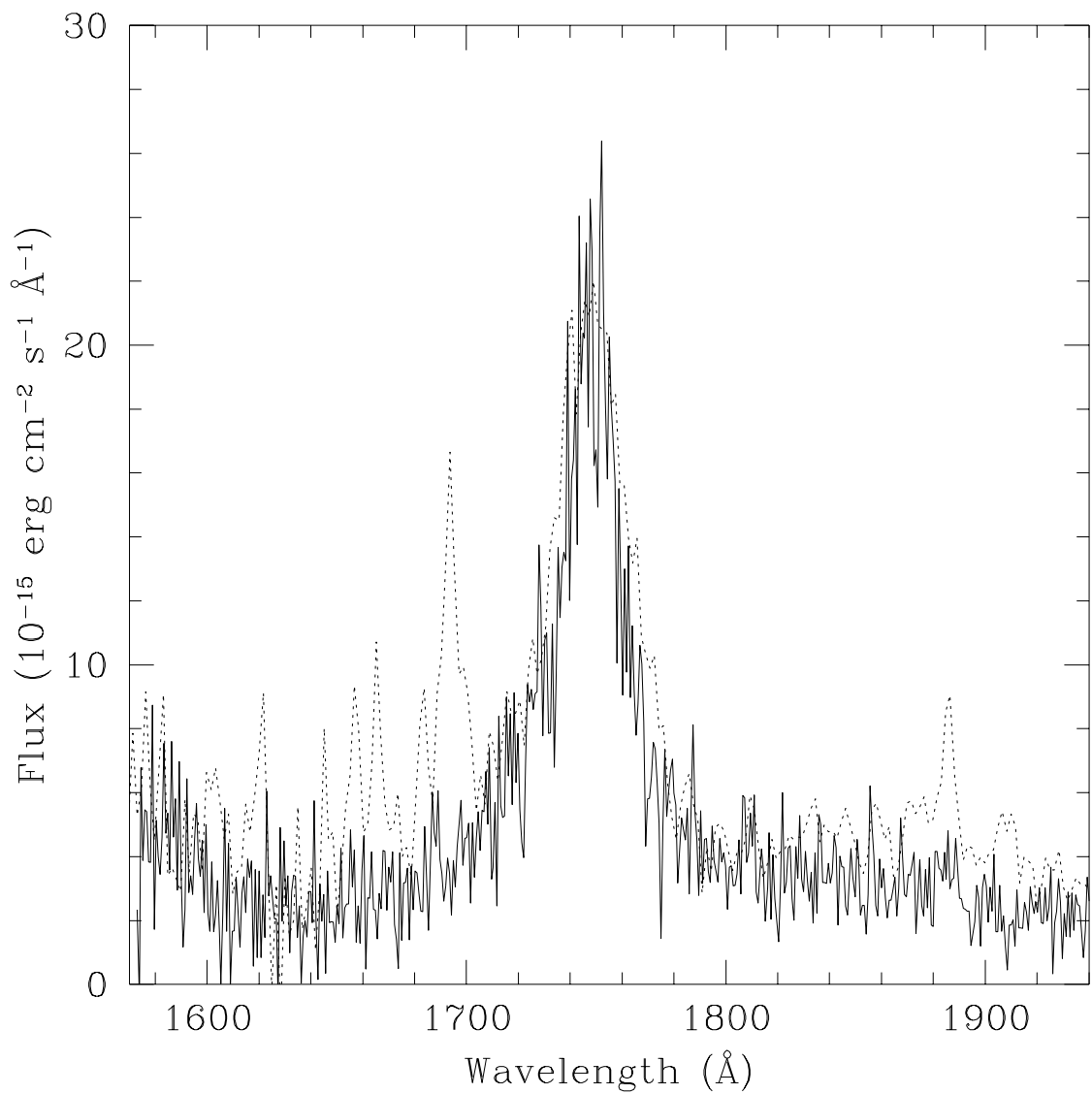}
\vfill
\caption{The CIV region of our HST FOS spectrum of
23 Aug 1996 (solid line), and of the highest SNR \IUE\, spectrum, SWP33030 from
03 Mar 1988 (dashed line). No evidence for BALs are seen in the HST
spectrum.  The overall line and continuum shape between the two
spectra are similar (no normalization has been applied). Flux spikes
at 1664\AA\, and 1694\AA\, in the \IUE\, spectrum are probably
spurious; one is a know artifact, and none of the other 4 UV spectra
reproduce these lines. 
}
\label{fspikes}
\end{figure}

The full spectrum, with a noise spectrum underlaid, is
shown is Figure~\ref{fullspec}. Given sufficient SNR, the spectral
resolution (R $\approx$ 2000) is adequate to measure narrow associated
lines (NALs) and to resolve some of the velocity structure of broad
lines.  The FOS spectrum shows no evidence for absorption either
narrow or broad.  For narrow lines, we used the software described in
Aldcroft et al. (1994), which iteratively fits for the quasar
continuum + emission line profile, and searches for narrow absorption
lines that are significant at 4-$\sigma$ confidence.  Within
6000\,km\,s$^{-1}$ of the quasar C\,IV emission line, the strongest
observed absorption line has a rest equivalent width of 1.7\,\AA,
while the detection limit is 2.3\,\AA.  For broad lines, it is more
difficult to establish formal detection criteria because the dominant
uncertainty is determination of the quasar continuum and emission line
profile.  In the case of \pg, the FOS spectrum is clearly consistent
with no BALs.  The features which lent an impression of BALs to the
\IUE\, spectra are revealed to be noise spikes, when the CIV region is
contrasted between SWP33030 and the FOS spectrum in
Figure~\ref{fspikes}.

\section{Variability}
\label{var}

There appears to be significant UV variability in \pg\, between the 
epochs of the observations presented here.  During the FOS and \IUE\,
observations (and from the \IUE\, observing logs, and the line-by-line
spectra) the QSO was always well centered in the aperture. 
We thus believe the changes in flux to be intrinsic to the QSO,
particularly because they are also accompanied by changes in 
emission line equivalent width.  

Is \pg\, unusually variable?  
For comparison to published results from |IUE\, spectra of a
large sample of AGN, we
calculated the standard deviation in continuum flux $f_{50}(\lambda)$
for two 50\AA\, bins centered at (rest) wavelengths of $\lambda=1450$
and 1625\AA, as outlined in Paltani \& Courvoisier (1994; PC94) (see
Table~\ref{tuv}).  The normalized 
variability index from $N$ epochs is calculated as

$$ \sigma_f(\lambda)= \frac{ \sqrt{
\frac{1}{N-1}\Sigma_{i=1}^{N}(f_{50,i}(\lambda)-\overline{f_{50}}(\lambda))^2}}
{\overline{f_{50}}(\lambda)}$$

\noindent and yields 0.813 and 0.681 at 1450 and 1625\AA,
respectively, using all 5 UV spectra of \pg.
At these wavelengths, the UV variability in RQ QSOs of similar
luminosity to \pg, as observed by \IUE\,
is typically about $34\pm 14\%$ (PC94).  \pg\, thus appears to be
unusually variable, similar to about 15\% of AGN that vary by more
than 50\% (i.e., $\sigma_f(\lambda)>0.5$; PC94).

 In addition to UV continuum variability, the CIV emission line flux
and \ew\, also changed significantly. There is no significant trend
with time for either line or continuum, and no correlation of
continuum level with line \ew\, (i.e., no Baldwin effect).

Is it possible that true BALs in \pg\, weakened or disappeared?
Absorption line variability has been seen both in narrower
associated absorbers, and in BALQSOs (Koratkar et al. 1996; Barlow
 1994).  However, no published record exists of any absorption
lines in BALQSOs completely
disappearing.  Rather, BAL variability is seen in about 15\% of
BALQSOs at the level of $20 - 40\%$ (Barlow
 1994).  Since BALs span a wide range in
velocity, most models of BAL cloud outflows require either an ensemble
of clouds along the line of sight, or winds blown off the surface of
the accretion disk (deKool \& Begelman 1995; Murray \& Chiang 1995).  
In either case, only small fractional changes in total absorption
column are expected.  Thus, since no BALs are presently seen in
this QSO, combined with the other evidence presented here,
we conclude that there are not now, and never have been
BALs in \pg.

\section{X-ray \& Gamma Ray Observations}

  \pg\, was strongly detected both by \ein\, and \ros\, in the 
soft X-ray bandpass, yielding consistent spectral fits typical 
of RQ QSOs.  Wilkes \& Elvis (1987) found a best fit powerlaw 
slope $\alpha_E=0.9^{+1.3}_{-0.6}$ ($\fnu\sim\nu^{-\alpha_E}$) with a
neutral (cold) absorption  
column of $\nh=1.2\times10^{21}$cm\mtwo with the \ein\, IPC
(0.3 - 3.5keV).  A \ros\, (0.1 - 2.4~keV) observation in 1992 January
yielded similar spectral results ($\alpha_E=1.2\pm0.15$;
deKool \& Meurs 1994), again with a cold absorbing column
($\nh=7\times10^{20}$ cm\mtwo) entirely consistent with the measured
21cm Galactic absorption. Assuming a warm (ionized) absorber, the best fit
spectrum yields $\nh<2\times10^{21}$cm\mtwo. Even this limit is far
below those derived from deep \ros\, observations of {\em bona fide}
BALQSOs presented in GM96.

 Earlier hard X-ray results from GINGA (1988 February) showed a very
flat spectrum between 2 and 20~keV ($\alpha_E\approx0.1$, Williams et
al. 1992), the hardest of all known AGN.  Since the flux at the low end
was well-matched to the later \ros\, fluxes, there was no
evidence for variability.

\pg\, varied at energies from 50 - 150~keV during (1994 September) OSSE
observations (Staubert, R. \& Maisack 1996). A powerlaw fit at these
energies is considerably steeper ($\alpha_E=2.2\pm0.5$) than for most
Seyfert galaxies ($\sim 1.2$).  If instead, the powerlaw slope is
fixed at the GINGA value  (whereby the normalization must have varied)
a good fit to the OSSE data requires an exponential cutoff of
e-folding energy $35\pm10$~keV, similar to many Seyfert galaxies.  

 Even though weakened soft X-ray emission suggestive of absorption is
observed in every BALQSO to date, the exact physical relation between
the UV and X-ray absorbers in BALQSOs is as yet unclear. It is thus
not obvious whether large changes in BALs would result in concomitant
changes in the emergent soft X-ray flux or spectral shape. However, we
would expect to see stronger changes in soft X-rays, if any, than at
the OSSE range. PG1416-129 never showed any soft X-ray absorption,
either before or after the \IUE\, observations.  We therefore view
substantial variation in the BALs in this QSO as an unlikely
explanation of the apparent discrepancy between the \IUE\, and \HST\,
spectra.

\section{Is \pg\, a BALQSO? }

It is clear from recent HST FOS spectra that \pg\, currently shows no
broad absorption lines.  Although there are some features resembling
BALs in the \IUE\, spectra, we argue for a
variety of reasons that \pg\, never exhibited BALs. 

From UV data, the \IUE\, spectra show spikes more consistent with recognized
artifacts, noise and/or defects such as cosmic rays than with 
the smooth broad troughs normally seen in BALQSOs.  These spikes vary
rather conspicuously with time in the \IUE\, spectra, but are not visible
in our new HST FOS spectrum.  Furthermore, based on observations
of other BALQSOs, it is not likely that there were true
BALs in \pg\, that have since vanished. 

All soft X-ray observations of \pg\, are consistent with no intrinsic
absorption either cold or warm.  Furthermore, no substantial variability
in either soft X-ray slope or normalization is observed, as might be
expected if indeed an absorber showed substantial changes 
in column or ionization state.

Finally, we consider optical emission lines. 
From a 1990 optical spectrum (Boroson \& Green 1992), \pg\, has
somewhat weak Fe\,II$\lambda 4500$ emission and average \roiii,
when compared to RQ QSOs of similar redshift.
By contrast, BALQSOs typically have strong iron emission
(Weymann et al. 1991), and small \roiii\, (at least for low-ionization
BALQSOs; Boroson \& Meyers 1992).  We also note from the \IUE\, spectra
of \pg\, that N\,V emission is virtually undetectable, even as an
asymmetry in the \lya\, profile, while Junkkarinen et al. (1987) 
suggest that N\,V in BALQSOs is marginally stronger than in 
non-BALQSOs.  As a caveat, some of these effects may depend on
redshift and/or luminosity.  Unfortunately, since \pg\, is the lowest
redshift (putative) BALQSO, no comparison samples of 
BALQSOs at similarly low redshift and luminosity exist.

Although the \HST\, spectrum alone shows definitively that \pg\,
has no UV BALs, we take the \IUE, \HST, X-ray and
optical spectral evidence together to show that \pg\, is not now, and
has never been a {\em bona fide} BALQSO.

\section{Discussion}
\label{discussion}

  GM96 compared their BALQSO sample to that of Laor et al. 1994 (L94),
who derived a mean \aox\, of $1.45\pm0.08$ for 10 RQ QSOs.  By
contrast, the BALQSOs in GM96 have a formal mean
$\overline{\aox}=2.17\pm0.1$ including \pg.  By removing \pg\, from
the BALQSO sample of GM96, we derive an overall sample mean \aox\, of
$2.24\pm0.08$.  Note that this formal "mean" is derived via survival
analysis, and includes only one detection, at $\aox=1.94$.

 However, \aox\, is known to depend on optical luminosity, and the
BALQSO sample and the L94 samples samples have significantly different
mean \lopt\, ($31.5\pm0.2$ vs. $30.5\pm0.1$ in the log, respectively,
with or without \pg).  From the general relationship between
\aox\, and \lopt\, (Wilkes et al. 1994, Avni et al. 1995), the
expected mean \aox\, values for the BALQSO and L94 samples are
$1.64\pm0.03$ and $1.53\pm0.02$, based only their optical 
luminosities.  A similarly significant
($\sim3\sigma$) difference between two samples of normal RQ QSOs
at these luminosities is predicted by the relationship derived in
Green et al. (1995), from a large, complete, optically selected
sample and different statistical techniques.  The observed difference
between the predicted mean value of \aox\, of normal RQ QSOs at
$\overline{\loglopt}=31.5$ and that observed for BALQSOs is thus
$\sim$0.6, about a $7\sigma$ disparity. 

 PG1416--129 might have been an outstanding exception that disproves the
rule, if indeed it were a true BALQSO.  An unabsorbed, X-ray bright
BALQSO would be a direct challenge
to some recent models that depict BALs as absorption of the nuclear
continuum by entrained winds off an accretion disk (Murray \& Chiang 1995). 
However, all {\em bona fide} BALQSOs have upon close examination
been X-ray quiet, suggesting strong absorption in soft X-rays.
Conversely, as we demonstrate here for one important case, it appears
that QSOs with a normal ratio of optical to soft X-ray flux \aox\, 
upon close examination will turn out {\em not} to be {\em bona fide}
BALQSOs.  Weak soft X-ray emission is a defining characteristic
of BALQSOs.  

The result that large \aox\, is observed in BALQSOs for
every case to date, together with the observation that intrinsic soft
X-ray absorption is rare in optically-selected QSOs (e.g., Laor et
al. 1997), suggests that the UV and soft X-ray absorbers have 
nearly the same covering factor and occupy the same solid angle as
seen from the QSO.  This  indicates that the UV and X-ray absorbers may
be closely related, if not identical.

 We note that true variability has been detected in UV BAL troughs: about
15\% of BALQSOs show variability in the residual UV intensity (rather
than in the velocity structure) at the $\sim20 - 40\%$ level, indicating
lower limits to column changes of about $10^{14}$cm\mtwo\, (Smith \&
Penston 1988, Barlow et al. 1992, Barlow 1994).  An ensuing change in
the measured soft X-ray absorption, however, has not yet been
observed.  Although 
we here rule out \pg\, from consideration as a BALQSO, a demonstration
of correlated variability between BALs and soft X-ray flux would
provide strong evidence that UV and soft X-ray absorbers are
physically associated in true BALQSOs. Such studies present a daunting
task for the current generation of X-ray telescopes, given the weak
soft X-ray fluxes of BALQSOs, but would  be feasible with the larger
effective areas of AXAF, XMM, or future high throughput X-ray
spectroscopy missions. These UV/X-ray variability studies could also
determine whether BAL variability is the result of a change 
in column density (e.g., due to motion of the absorber
along the line of sight), or in ionization.

Our thanks to the referee, Fred Hamann, for his comments
and insight. PJG, TLA, and SM gratefully acknowledge support provided by NASA
Grant GO-06528.01-95A, and PJG the support provided by NASA
through Grant HF-1032.01-92A, both awarded by the Space Telescope Science 
Institute, which is operated by the Association of Universities for
Research in Astronomy, Inc., under NASA contract NAS5-26555. 
PJG and TLA also acknowledge support through NASA Contract NAS8-39073
(ASC), and SM was also supported by NASA grant NAGW-4490 (LTSA).


\begin{deluxetable}{lrrrrr}
\small
\tablewidth{33pc} 
\tablecaption{Basic Data on UV Spectra of \pg 
\label{tuv}} 
\tablehead{
\colhead{Image} &
\colhead{Date}  &
\colhead{Exposure} &
\colhead{$f_{1450}$ \tablenotemark{a}} &
\colhead{$f_{1625}$ \tablenotemark{a}} &
\colhead{\wciv\tablenotemark{b} } \\
\colhead{} &
\colhead{}  &
\colhead{(s)} &
\multicolumn{2}{c}{$10^{-15}$ \fnucgs } &
\colhead{(\AA ~ rest) } \\
}
\tablecolumns{6}
\phm{STRING}  
\startdata
{\em IUE:} &  & & & & \nl
SWP08916 & 04 May 1980 & 8700 & 6.58 $\pm$3.49 & 7.20 $\pm$1.56 & 158 $\pm$14 \nl
SWP16763 & 14 Apr 1982 & 1680 & 13.2 $\pm$12.8 & 11.8 $\pm$5.55 & $<108$\phm{ $\pm$18} \nl
SWP33030 & 03 Mar 1988 & 24900 & 4.53 $\pm$2.62 & 4.68 $\pm$0.64 & 126 $\pm$18 \nl
SWP45019 & 27 Jun 1988 & 24900 & 1.62 $\pm$1.37 & 1.74 $\pm$0.51 & 96 $\pm$18 \nl
{\em HST:} &  & & & & \nl
Y3DB0103T & 23 Aug 1996  & 940 &  2.49 $\pm$1.77 & 3.39 $\pm$1.46 &
182 $\pm$14 \nl

\enddata 
\tablenotetext{a}{Mean (observed frame) continuum fluxes, measured at
the (rest) 
wavelength indicated, in bins 50\AA\, (also rest).}
\tablenotetext{b}{Emission line equivalent width.}
\end{deluxetable} 

\clearpage

\end{document}